\documentclass[a4paper,sort&compress,5p]{elsarticle}

\usepackage{multirow,amsmath}

\usepackage{lineno,hyperref}
\modulolinenumbers[5]

\journal{Physics Letters B}

\begin{document}

\begin{frontmatter}

\title{Observational constraints on the unified dark matter and dark energy model based on the quark bag model}


    \author[ipn]{Ariadna Montiel }
    \ead{amontiel@fis.cinvestav.mx}
    \author[upv]{Vincenzo Salzano}
    \ead{vincenzo.salzano@ehu.es}
      \author[upv]{Ruth Lazkoz}
    \ead{ruth.lazkoz@ehu.es}
    
\address[ipn]{Departamento de F\'{\i}sica, Centro de Investigaci\'on y de Estudios Avanzados del I. P. N., Apartado Postal 14-740, 07000 M\'exico D.F., Mexico.} 
\address[upv]{Departamento de F\'{\i}sica Te\'orica e Historia de la Ciencia, Universidad del Pa\'{\i}s Vasco (UPV/EHU), Apdo. 644, E-48080, Bilbao, Spain}




\begin{abstract}
In this work we investigate if a small fraction of quarks and gluons, which escaped hadronization and survived as a uniformly spread perfect fluid, can play 
the role of both dark matter and dark energy. This fluid, as developed in \citep{Brilenkov}, is characterized by two main parameters: 
$\beta$, related to the amount of quarks and gluons which act as dark matter; and $\gamma$, acting as the cosmological constant. 
We explore the feasibility of this model at cosmological scales using data from type Ia Supernovae (SNeIa), Long Gamma-Ray Bursts (LGRB) 
and direct observational Hubble data. We find that: (i) in general, $\beta$ cannot be constrained  by SNeIa data nor by LGRB or H(z) data; (ii) 
$\gamma$ can be constrained quite well  by all three data sets, contributing with $\approx78\%$ to the energy-matter content; (iii) when a strong prior on (only) 
baryonic matter is assumed, the two parameters of the model are constrained successfully. 
\end{abstract}


\end{frontmatter}


\section{Introduction}
A huge amount of high-quality observational data collected so far has made the acceleration of the universe an indisputable fact 
\citep{Riess:1998cb,Perlmutter:1998np,Knop:2003iy, Riess:2004nr,Astier:2005qq,Spergel:2003cb,Spergel:2006hy,Tegmark:2003ud,PlanckXVI}. 
Such unexpected behaviour has been commonly attributed to an unknown entity acting as a counter-gravitating fluid, dark energy (DE),
and has motivated the bloom of an impressive amount of cosmological models which may be able to elucidate its nature. So far, the so called $\Lambda$CDM 
model is the most accepted cosmological model, and it is based on the well known cosmological constant; however, it still suffers from theoretical drawbacks 
that make it difficult to reach a conclusive consensus.

Many theoretical proposals can be found which attempt  to throw some light on the cosmic acceleration mystery, either trying to address its very origin, 
or (more modestly) attempting at a compelling description of the recent history of our accelerated universe. Some proposals are  based on scalar fields, 
either canonical such as quintessence \citep{Wetterich1988,Peebles:1987}, or with weirder features, such as k-essence \citep{Armendariz2000} or phantom 
\citep{Caldwell2003} models. Others have an extra-dimensional spirit
and  invoke braneworlds \citep{LisaRandall1999,LisaRandall1999b}.  

Dark matter (DM) is the other main, yet unknown, component of the Universe, and it is necessary to produce enough gravitational attraction on certain scales
crucial to structure formation. 
Some of the proposals for the description of accelerated cosmologies rely on (phenomenological) unified pictures (so called unified dark matter models)
where a unique exotic fluid accounts for the whole dark sector composed by DE and DM. 
If we specifically refer to  unified dark matter models, then let us remind that most of them resort to the generalized Chaplygin gas (GCG) 
\citep{Kamenshchik2001265,Bilic200217,Bento2002}, but one can find other (also phenomenological) proposals as those in \citep{Bertacca2010,Bertacca2011}.

The list of (accelerated) scenarios can be completed with many other cases. But if  we use the popularity criterion 
among those additional proposals, then modifications to the General Relativity Lagrangian stand out  \citep{Capozziello:2003tk,Capozziello:2003gx,Carroll:2003wy,Jain:2010ka}. 
Nevertheless see \citep{Sahni2000,Peebles2003,Padmanabhan2003,Sahni:2005,Copeland,Frieman2008,Li:2011sd,Mortonson:2013zfa} 
for  reviews on DE models, which provide a wide perspective on the topic of current cosmic acceleration in general.

In general we have a vast collection of set-ups which are quite different in their underlying physics, and although many of them (including the {\it concordance} $\Lambda$CDM model)
have a great compliance with observational data, none of them is full proof.

On the other hand, the proposal by \citep{Brilenkov}, which can be considered as part of the stream of unified dark matter models, has been suggested to explain the nature of
DM and the present cosmic acceleration. Such suggestion arises from the hypothesis that a small part of quarks and gluons did not yield to hadronization, and  resisted either as
isolated aggregates of quark-gluon \textit{nuggets} (QNs) or as a perfect fluid in the form of a quark-gluon plasma (QGP) (uniformly spread on cosmological scales). There have been several
works scrutinizing and supporting this guess  \citep{Witten:1984rs,Applegate:1985qt,Farhi:1984qu,Chandra:1999tr}, and the idea followed that the QNs could be a good candidate for DM. 
In fact, a recent perturbative analysis \citep{Brilenkov2}, reached the conclusion that compatibility with observations was possible (for a mechanical perspective on this topic see 
\citep{Eingorn:2012jm,Eingorn:2012dg,Eingorn:2013faa}).

In contrast, the QGP perfect fluid has not gathered the same interest. A recent work \citep{Rahaman:2012tu} explored the possibility that  the QGP fluid acted as DM in galactic halos  
concluding that the corresponding rotation curves were reasonable. At the cosmological level, a QGP fluid  was first considered to mimic DM
in \citep{Brilenkov}. 


Clearly, the theoretical perspective makes the quark bag an attractive one, as it opens the door to an answer to the nature of the two dark components 
of the universe without resorting
to exotic physics. 

In previous reference the compliance with observations was carried out in an inverse approach, assuming facts hinted by observations the necessary properties 
of the quark models were derived. But it is absolutely mandatory to work reversely, that is, to assume the model and then to contrast its theoretical predictions 
with the observational data. This has to be done in an statistically proper way, beyond quantitative sketches. A thorough study will 
allow to ascertain whether the model 
is worth exploring further. This is precisely the goal of  this work: to establish the viability of the quark bag proposal, which attempts to explain DE and DM in a unified fashion. 
In order to do that, we perform a standard statistical analysis by using the following astrophysical probes: type Ia Supernovae (SNeIa), Long Gamma-Ray Bursts (GRBs) 
and observational Hubble data. 
In the next section, we describe briefly the main conclusions of each scenario sketched in \citep{Brilenkov}. 
In Section III, we present the observational data samples used in our analysis and finally, in Section IV, we describe and discuss our findings.

\section{Cosmological scenarios from QGP}

In this section we shall focus  on the cosmological consequences of the two theoretical scenarios proposed in \citep{Brilenkov}, i.e. QNs and QGP. These two scenarios
begin to be valid at  the matter dominated era and then remain so forever. 
Even though these two set-ups are based on the quark bag equation of state, we will show that each of them has different cosmological implications, see \citep{Brilenkov} for further details.
Besides, following the same reference, we assume the quark bag fluid is accompanied by a a cosmological constant.


Notice also that throughout this analysis we have considered $\Omega_K=0$ as has been recently confirmed by \citep{PlanckXVI}.

\begin{table*}[!htp]
\caption{Summary of the cosmological scenarios, Quark Nuggets and Quark-Gluon-Plasma-like perfect fluid, and their cosmological \hbox{implications.}}
\renewcommand{\arraystretch}{2.}
\begin{tabular}{l @{\extracolsep{10pt}} l @{\extracolsep{15pt}} l }
\hline
Models & Parameters & Conclusion  \\ \hline
\vspace{2mm}
\multirow{2}{*}{QNs(I)} & $\Omega_{\Lambda}=0$, $\beta\neq0$ & \small{Unable to explain current cosmological acceleration: $q_0<0$ only for non-physical $\beta <0$} \\
 & $\Omega_{\Lambda}\neq 0$, $\beta=0$ & \small{QNs are candidates for DM but not for DE}, $\Omega^{\rm eff}_{m}=\gamma^{4/3}+\Omega_m$ \\ \hline
 \vspace{2mm}
\multirow{2}{*}{QNs(II)} & $\Omega_{\Lambda}=0$ &\small{Impossible to explain current cosmological acceleration; $\Omega_{\Lambda}$ forcefully needed} \\
 & $\Omega_{\Lambda}\neq 0$ & \small{QNs are candidates  for DM only}, $\Omega^{\rm eff}_{m}=\gamma+\Omega_m$ \\ \hline
 \vspace{2mm}
\multirow{2}{*}{QGP(I)} & $\Omega_{\Lambda}=0$, $\beta=0$ &\small{$\Lambda$CDM model reproduced: $\Omega^{\rm eff}_{\Lambda}=\gamma^{4/3}$ and $\Omega^{\rm eff}_{m}=\Omega_m$} \\
 & $\Omega_{\Lambda}= 0$, $\beta\neq0$ & \small{QGP possible candidate to both dark matter and dark energy}  \\ \hline
\vspace{2mm}
QGP(II) & $\Omega_{\Lambda}=0$, $\beta\neq0$ & \small{$\Lambda$CDM model reproduced: $\Omega_{\Lambda,qgp}=\gamma$, $\Omega^{\rm eff}_{m}=\Omega_m$ and $\Omega^{\rm eff}_{rad}=\beta$} \\ \hline
\end{tabular}
\label{Table:Models}
\end{table*}

\subsection{Quark Nuggets (I)}

This scenario stems from  the modification of the quark bag equation of state (EoS) suggested in \citep{Kallman1984363}. From such
modified EoS one gets the following Hubble function:
\begin{equation}
\mkern-10mu \frac{H^2}{H_0^2}=\left[ \beta \left(\frac{a_0}{a}\right)^3+ \gamma\left(\frac{a_0}{a}\right)^{9/4}\right]^{4/3}+\Omega_m \left(\frac{a_0}{a}\right)^3+\Omega_{\Lambda},~
\label{Eq:HubbleQNI}
\end{equation}
where, in principle, the $\Omega_m$ term corresponds to the usual matter content at present (baryons+DM), and $\Omega_{\Lambda}$ to the cosmological constant. Notice that the term inside the brackets behaves 
at early times $(a \ll a_0)$ like radiation, $\sim \beta a^{-4}$, and at later times $(a \gg a_0)$ like matter, $\sim \gamma a^{-3}$.

Concerning the cosmological implications of this model, not many conclusions can be drawn. First, 
QNs alone cannot drive cosmic acceleration, the cosmological term is necessary for that.
On the contrary, when $\Omega_{\Lambda}=0$ is assumed, the cosmic acceleration 
is achieved only if $\beta<0$, which leads to inconsistencies in the model because $\beta$ is positive definite, see \citep{Brilenkov} for further details. 

On the other hand, assuming the presence of the cosmological constant, and by considering $\beta=0$ (or the weaker condition $\beta \ll \gamma$, which is quite realistic, 
if the term proportional to $\beta$ acts as radiation), one easily obtains:
\begin{equation}
\Omega^{\rm eff}_{m} = \gamma^{4/3} + \Omega_{m} \; .
\end{equation}
In this case, $\Omega_{m}$ could play only the role of baryonic matter, and $\gamma^{4/3}$ that of DM. 
Unfortunately it is quite clear that the use of observational data at the background level will not allow to distinguish between this model and the standard  literature results. Phenomenologically all remains very much the same, only the theoretical interpretation about the origin of the model is new.

A more interesting case arises when one chooses $\beta \neq 0$ and then wonders whether $\gamma$ assumes values compliant 
with the proposed assumption of quarks acting like dark matter. Let us recall that the weaker condition $\beta \ll \gamma$ could realistically hold,  if the term containing $\beta$ acted as radiation. But the role of $\gamma$ as a possible contribution to 
DM
has to be verified. We will study this case in the next sections.

\subsection{Quark Nuggets (II)}

The cosmological implications of this model follow from  the assumption of the original quark-bag EoS, see Section 3.2 of \citep{Brilenkov}. The Hubble function is given by:
\begin{equation}
\frac{H^2}{H_0^2}= \beta \left(\frac{a_0}{a}\right)^4+ \gamma\left(\frac{a_0}{a}\right)^{3}+\Omega_m \left(\frac{a_0}{a}\right)^3+\Omega_{\Lambda} \; .
\label{Eq:HubbleQNb}
\end{equation}
Clearly, the main difference with model I, is the absence of any interaction between the $\beta$ and the $\gamma$ terms. As can be seen from Eq. (\ref{Eq:HubbleQNb}), this scenario contains standard (i.e. isolated) radiation and matter components, but their origin is from the thermodynamical properties 
of QNs. Obviously, this model goes to the
the $\Lambda$CDM case when one assumes that the effective matter content is $\Omega^{\rm eff}_{M}=\gamma+\Omega_m$, where $\Omega_{m}$ could play the role of only baryonic matter. On the other hand, it is not possible to explain current cosmic acceleration without the cosmological constant, 
whereas the QNs may be candidates  for DM only, but again, the situation is observationally indistinguishable from other classical interpretations.

\subsection{Quark-Gluon-Plasma-like perfect fluid (I)}

This model assumes that the perfect fluid composed by a quark-gluon plasma has thermodynamical properties derived from the modified quark-bag EoS proposed by \citep{Kallman1984363}. The Hubble function, see Section 4.1 of \citep{Brilenkov} for further details, is given by:
\begin{equation}
\frac{H^2}{H_0^2}=\left[ \beta \left( \frac{a_0}{a} \right) ^3 + \gamma \right]^{4/3}+ \Omega_m \left(\frac{a_0}{a} \right)^3+\Omega_{\Lambda}.
\label{Eq:HubbleQGPI}
\end{equation}
In the particular case of $\beta=0$, the $\Lambda$CDM model is clearly restored and the term containing $\gamma^{4/3}$ might play the role of the cosmological constant. Thus, in principle one could set $\Omega_{\Lambda}=0$, and a satisfactory fit to cosmological data would be possible, but quarks would not contribute to DM.

In \citep{Brilenkov}, the more interesting $\beta \neq 0$, $\Omega_{\Lambda}=0$  case is considered, upon the hypothesis  that 
$\Omega_m$  \textit{should} correspond only to baryonic matter, while the $\beta$ and $\gamma$ parameters \textit{might} account for the nature of DM and DE, respectively. We will explore the feasibility of this model in more detail in the following section.

\subsection{Quark-Gluon Plasma like perfect fluid (II)}

This is our last scenario, with the Hubble function given by:
\begin{equation}
\frac{H^2}{H_0^2}= \beta \left( \frac{a_0}{a} \right) ^4 + \gamma + \Omega_m \left(\frac{a_0}{a} \right)^3+\Omega_{\Lambda}.
\label{Eq:HubbleQGPb}
\end{equation}
The $\Lambda$CDM model is recovered even if one takes $\Omega_{\Lambda}=0$. In this situation, the $\gamma$ parameter plays the role of the cosmological constant, and the term which includes $\beta$ acts as radiation, but it would be 
impossible to disentangle the contribution of quarks to $\Omega_{m}$ from classical DM.

\subsection{Model summary}

For a summary of all these cosmological scenarios and their principal cosmological consequences, see Table \ref{Table:Models}.

As we have mentioned above, the scenarios called QN(I) (with $\beta \neq 0$) and QGP(I) seem to be the most interesting toward a confrontation with observational data, since the QN(I) scenario presumes quarks could act like DM, while the QGP(I) scenario could explain the acceleration of 
the Universe in the absence of a cosmological constant and at the same time could account for DM. So, hereafter, we are going to focus on these scenarios and their 
Hubble function, given by  Eq. (\ref{Eq:HubbleQNI}) by assuming $\beta \neq 0$ and Eq. (\ref{Eq:HubbleQGPI}) with $\Omega_{\Lambda}=0$, respectively.

As customary (and setting $a_0=1$), consistency of Eqs.~(\ref{Eq:HubbleQNI}) and (\ref{Eq:HubbleQGPI}) translates into the conditions
\begin{align*}
\left(\beta+\gamma \right)^{4/3} + \Omega_m + \Omega_{\Lambda}=1& \quad \mathrm{for} \quad \mathrm{QN(I)} ,\nonumber \\
\left(\beta+\gamma \right)^{4/3} + \Omega_m=1 &\quad \mathrm{for}\quad \mathrm{QGP(I)}.
\end{align*}
Thus the dimensionality of our statistical analysis can be reduced to only three ($\Omega_{m},\beta,\gamma$) and two ($\beta,\gamma$) free parameters, respectively.

The deceleration parameter for the QN(I) and QGP(I) scenarios are given by
\begin{align}
 \label{Eq:q}
q(z)&= \left( \frac{H_0}{H} \right)^2 \left\{\left[\beta \left(\frac{a_0}{a} \right)^3 + 
\gamma \left(\frac{a_0}{a} \right)^{9/4} \right]^{4/3}+\frac{\Omega_m}{2}  \left( \frac{a_0}{a} \right)^3  \right. \nonumber \qquad\\
& \left. -\frac{\gamma}{2} \left[\beta \left(\frac{a_0}{a} \right)^{39/4} + \gamma \left(\frac{a_0}{a} \right)^{9} \right]^{1/3}-\Omega_{\Lambda}\right\} \quad \mathrm{for} \quad \mathrm{QN(I)} \nonumber \\
q(z)&= \left( \frac{H_0}{H} \right)^2 \left\{ \left[ \beta \left(\frac{a_0}{a} \right)^3 + \gamma \right]^{4/3} +\frac{\Omega_m}{2}  \left( \frac{a_0}{a} \right)^3 \right. \nonumber \\
& \left. - 2\gamma \left[\beta \left(\frac{a_0}{a} \right)^3 + \gamma \right]^{1/3}   \right\} \quad \mathrm{for} \quad \mathrm{QGP(I)}
\end{align}
so that, at $a=a_0(=1)$ Eq. (\ref{Eq:q}) takes the form:
\begin{align*}
q_0&=\left(\beta+\gamma \right)^{4/3}+\frac{\Omega_m}{2} -\frac{\gamma}{2} \left(\beta+\gamma \right)^{1/3} -\Omega_{\Lambda} \quad \mathrm{for} \quad \mathrm{QN(I)}, \nonumber \\
q_0&= \frac{\Omega_m	}{2} +\left(\beta+\gamma \right)^{4/3} -2\gamma \left(\beta+\gamma \right)^{1/3} \quad \mathrm{for} \quad \mathrm{QGP(I)}.
\label{Eq:q0}
\end{align*}

\section{Observational data }

\subsection{Type Ia Supernovae}

We have used the Union 2.1 compilation released by \cite{Union21} as our supernovae data set. This compilation consists of 580 SneIa, distributed over the redshift interval $0.015 < z < 1.4$, 
and is one of the largest and spectroscopically confirmed samples. For statistical tests of the Union 2.1 SNeIa sample one uses the definition of the 
distance modulus:
\begin{equation}
\mu(z_j)= 5 \log_{10} [d_L(z_j, \theta_i) ] + \mu_0,
\label{Eq:mu}
\end{equation}
with $\mu_0=42.38-5\log_{10} h$, with $h$ being the dimensionless Hubble constant, and $d_L(z_j,\theta_i)$ being the Hubble free luminosity distance defined as
\begin{equation}
d_L(z,\theta_i)= (1+z)\int^z_0  \frac{dz'}{E(z',\theta_i)},
\end{equation}
where $E(z,\theta_i)=H(z,\theta_i)/H_0$ and the $\theta_i$ stand for  the vectors of parameters of the model.

The $\chi^2$ function for the SNeIa data is
\begin{equation}
\chi^2_\mu (\mu_0, \theta_i)= \sum^{580}_{j=1} \frac{(\mu(z_j; \mu_0, \theta_i)-\mu_{obs}(z_j))^2}{\sigma^2_{\mu}(z_j)},
\label{Eq:ChiMu}
\end{equation}
where the $\sigma_{\mu}(z_j)$ represent the uncertainties on the distance modulus for each supernova. The parameter $\mu_0$ in Eq. (\ref{Eq:mu}) has to be marginalized over, as it  is a nuisance parameter  (the reason being it encodes the Hubble parameter and the absolute magnitude $M$) . We have found it convenient 
to work with a reformulation of Eq. (\ref{Eq:ChiMu}) suggested by \cite{Pietro03,Nesseris05}, which follows from minimizing the $\chi^2$ function with respect to $ \mu_0$. Then, one can rewrite Eq. (\ref{Eq:ChiMu}) as
\begin{equation}
\chi^2_{SN} (\theta)= c_1 - 2c_2 \mu_0 + c_3 \mu^2_0,
\end{equation}
with
\begin{align*}
c_1&=\sum^{580}_{j=1} \frac{(\mu(z_j; \mu_0=0,\theta_i)-\mu_{obs}(z_j))^2}{\sigma^2_{\mu}(z_j)}, \nonumber \\
c_2&=\sum^{580}_{j=1} \frac{(\mu(z_j; \mu_0=0,\theta_i)-\mu_{obs}(z_j))}{\sigma^2_{\mu}(z_j)}, \\
c_3&=\sum^{580}_{j=1} \frac{1}{\sigma^2_{\mu}(z_j)}. \nonumber \\
\end{align*}
Upon minimization over $\mu_0$ one gets $\mu_0=c_2/c_3$. Finally, the $\chi^2$ function reads
\begin{equation}
\tilde{\chi}_{SN} (\theta_i)= c_1 - \frac{c^2_2}{c_3}.
\label{Eq:chiSN}
\end{equation}
Since $\tilde{\chi}^2_{SN}=\chi^2_{SN}(\mu_0=0,\theta_i)$ (up to a constant), we have minimized $\tilde{\chi}^2_{SN}$ instead of the usual expression.

\subsection{Long Gamma-Ray Bursts}

Gamma-Ray Bursts (GRBs) are astrophysical phenomena for which typically one can get observational data at higher redshifts than for SNeIa. 
Therefore, GRBs data offer tracks to investigate cosmological models at these high redshifts. Unfortunately, from a strict point of view, GRBs are not standard candles like SNeIa, and thus
an appropriate calibration is necessary to regard them as reliable distance indicators. Their use in the cosmological battlefield has motivated many empirical luminosity correlations, although the lack of low redshift GRBs data typically make   calibrations cosmological model dependent. This difficulty is known as the circularity problem, 
and several efforts to do away with it have been made, see for example \cite{Kodama08,Liang08,Wei09,Wei10,Wang08,Cardone09}. To make matters more complicated, uncertainties on the observable quantities of
GRBs are much larger 
than for SNeIa,
letting alone the fact that there is not so far a good understanding of their source mechanism. These problems favour an active controversy about the use of GRBs for cosmological purposes, see, e.g., \cite{Cuesta,Liang10,Freitas11,Graziani,Collazzi,Butler,Sha,Butler10}, therefore the 
choice of a good GRB sample is essential.

Recently, in \cite{Yonetoku12},  a set of 9 Long Gamma-Ray Bursts (LGRBs) in the redshift range $1.547 \leq z \leq  3.57$ has been calibrated through the Type I Fundamental Plane. This is defined by the correlation between the spectral peak energy $E_p$, the peak luminosity $L_p$, and the luminosity time $T_L\equiv E_{iso}/L_p$, where $E_{iso}$ is the isotropic energy. This calibration is one of the several 
proposals to calibrate GRBs in an cosmology-independent way. The fact that a control of systematic errors has been carried out to calibrate these 9 LGRBs \cite{Yonetoku12} makes this compilation a very compelling one; thus we have included it in our analysis.

The $\chi^2$ function for the GRB data is defined by
\begin{equation}
\chi^2_{LGRBs}(\theta_i )= \sum_{j=1}^{9} \frac{(\mu_{th}(z_j, \theta_i) - \mu_{\rm obs}(z_j))^2}{\sigma^2_{\mu}(z_j)},
\label{Eq:CLGRB}
\end{equation}
where $\mu_{th}(z_j,\theta_i)= 5 \log_{10} [d_L(z_j, \theta_i) /{\rm Mpc}] + 25$ and the $\sigma_{\mu}(z_j)$ are the measurement errors on the distance modulus. 
We have also fixed $H_0$ as $70~{\rm km} {\rm s}^{-1} {\rm Mpc}^{-1}$ \citep{Yonetoku12}, because this value was used to derive the distance modulus values, and leaving it free may induce an unwanted cosmological model bias.

\subsection{Hubble parameter}

The differential evolution of early-type galaxies with passive evolution provides direct measurements of the Hubble parameter, $H(z)$. An updated compilation of such data
was presented in \cite{Jimenez12}, whereas older data can be found in \cite{Jimenez02}. 
As this data set avoids one level of integration  with respect to other observational tools, like SNeIa, GRBs (and angular/angle-averaged BAO), the
well-known and somewhat unwanted smearing effect which plagues these other observables is not so severe.
This property favours the use of these data set for useful consistency checks or tighter constraints on models.

In this work we adopt the 18 data points in the redshift range $0.09\leq z \leq1.75$ reported in \cite{Jimenez12}, and we use them to estimate the model parameters by minimizing the quantity
\begin{equation}
\chi^2_{H} (H_0,\theta_i)= \sum^{18}_{j=1} \frac{\left[ H_{th}(z_j,\theta_i)-H_{obs}(z_j)\right]^2}{\sigma^2_{H_{obs}}(z_j)},
\label{Eq:COHD}
\end{equation}
where $H_0= 100 \; h \; {\rm km} {\rm s}^{-1} {\rm Mpc}^{-1}$ will be fixed at $H_0=67.3~{\rm km} {\rm s}^{-1} {\rm Mpc}^{-1}$ \citep{PlanckXVI} and $\sigma^2_H$ are the measurement variances.

\section{Results and Discussion}

\begin{figure}
  \centering
  \includegraphics[width=0.45\textwidth]{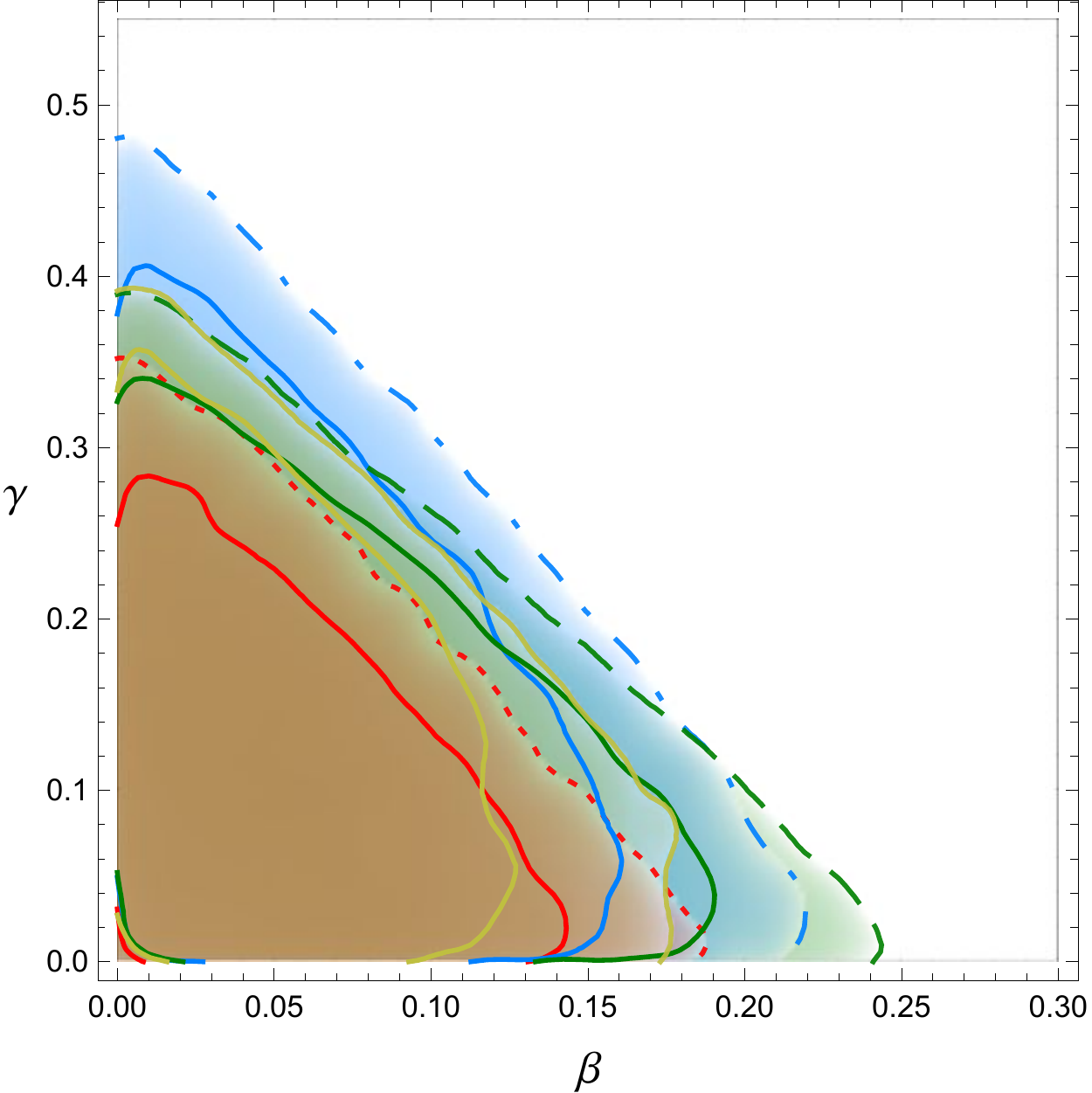}
  \hskip .8cm
    \includegraphics[width=0.45\textwidth]{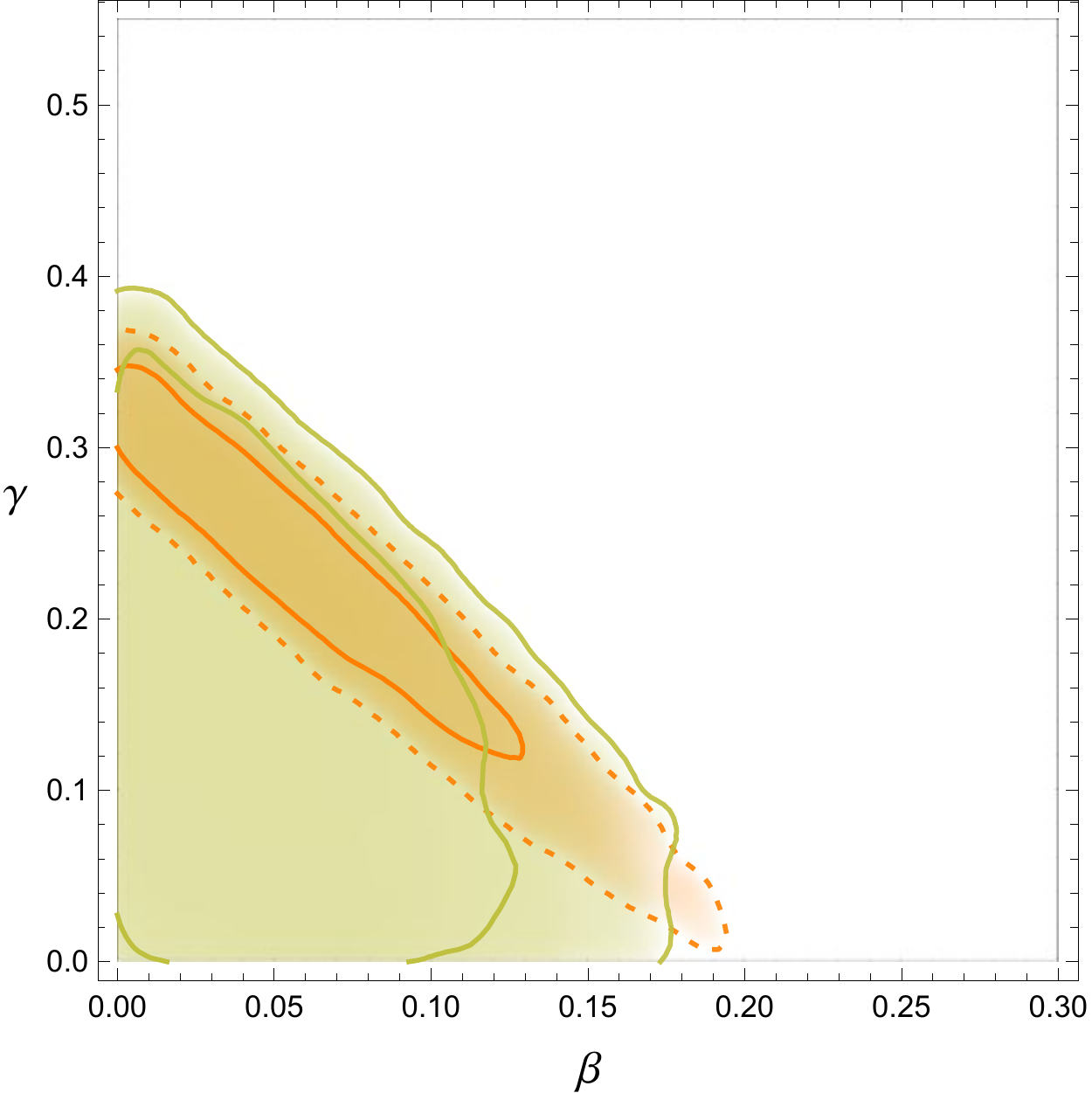}
  \caption{\label{Fig:ContoursQNI} Confidence regions in the $(\beta,\gamma)$ plane for the QNs(I) model. (\textit{First panel.}) The contours correspond to: LGRBs (shaded region in red; 
  the $1\sigma$ and $2\sigma$ confidence levels correspond to the solid line and to the dotted line, respectively), 
  SNeIa (shaded region in green; the $1\sigma$ and $2\sigma$ confidence levels correspond to the solid line and to the dashed line, respectively), H(z) data (shaded region in blue; 
  the $1\sigma$ and $2\sigma$ confidence levels correspond to the solid line and to the dot-dashed line, respectively) and 
  LGRBs+SNeIa+H(z) data (shaded region in yellow; the $1\sigma$ and $2\sigma$ confidence levels are drawn in solid lines). (\textit{Second Panel.}) The contours correspond to 
  $1\sigma$-$2\sigma$ confidence levels using LGRBs+SNeIa+H(z) data. The shaded region in orange shows the confidence region that results from the $\chi^2$ analysis by assuming a Gaussian prior on $\Omega_m$ from \citep{PlanckXVI}, while the shaded region in 
yellow is obtained without assuming a prior knowledge on $\Omega_m$.  }
\end{figure}

\begin{figure}
  \centering
  \includegraphics[width=0.45\textwidth]{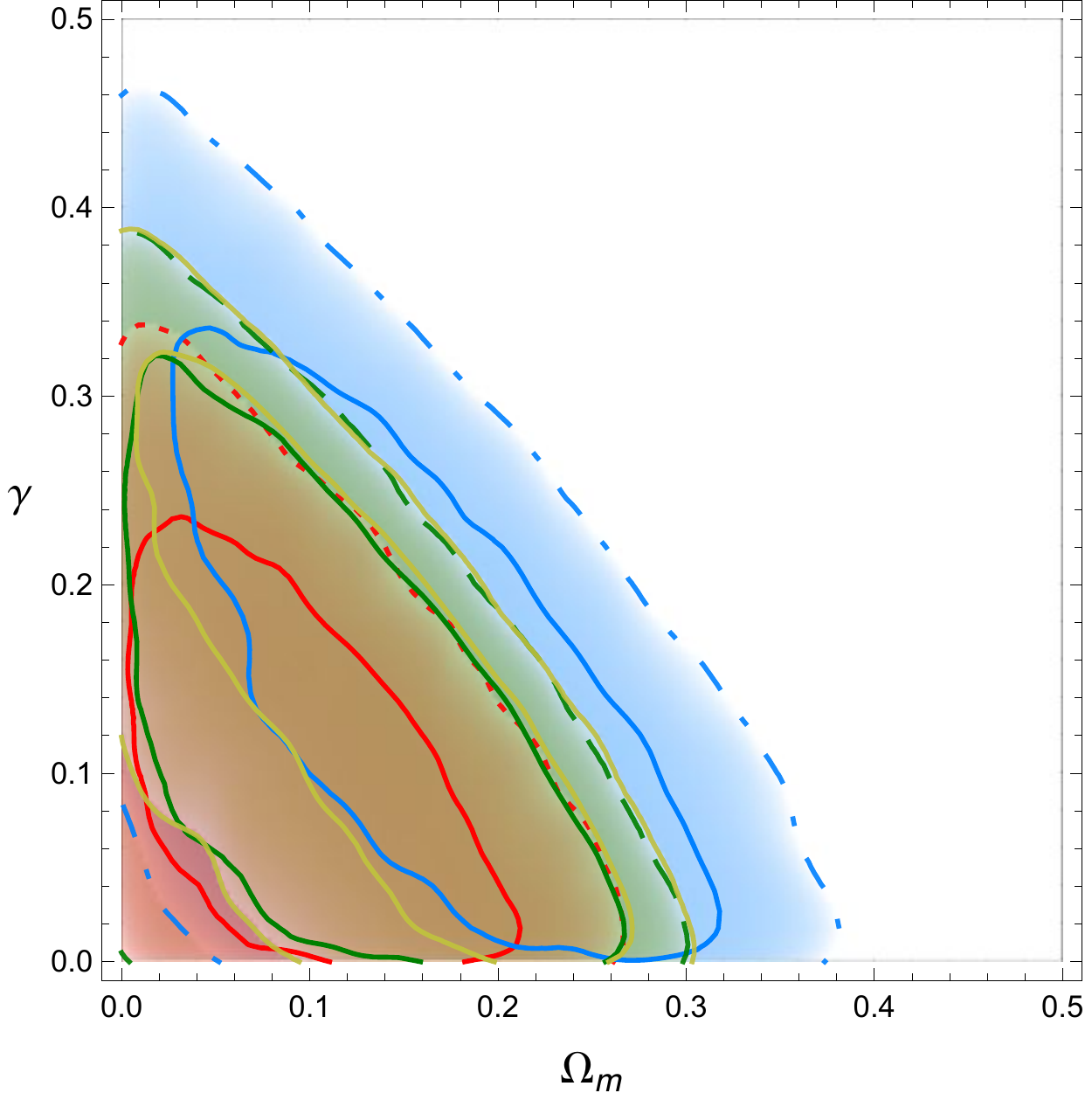}
    \caption{\label{Fig:ContoursQNIOmG} Confidence regions in the $(\Omega_m,\gamma)$ plane for the QNs(I) model. The contours correspond to $1\sigma$-$2\sigma$ confidence levels using LGRBs, SNeIa, H(z) data and LGRBs+SneIa+H(z) data. The code of colours is the same as on the 
    first panel of Figure 
    \ref{Fig:ContoursQNI}. }
\end{figure}

\begin{figure}
  \centering
  \includegraphics[width=0.45\textwidth]{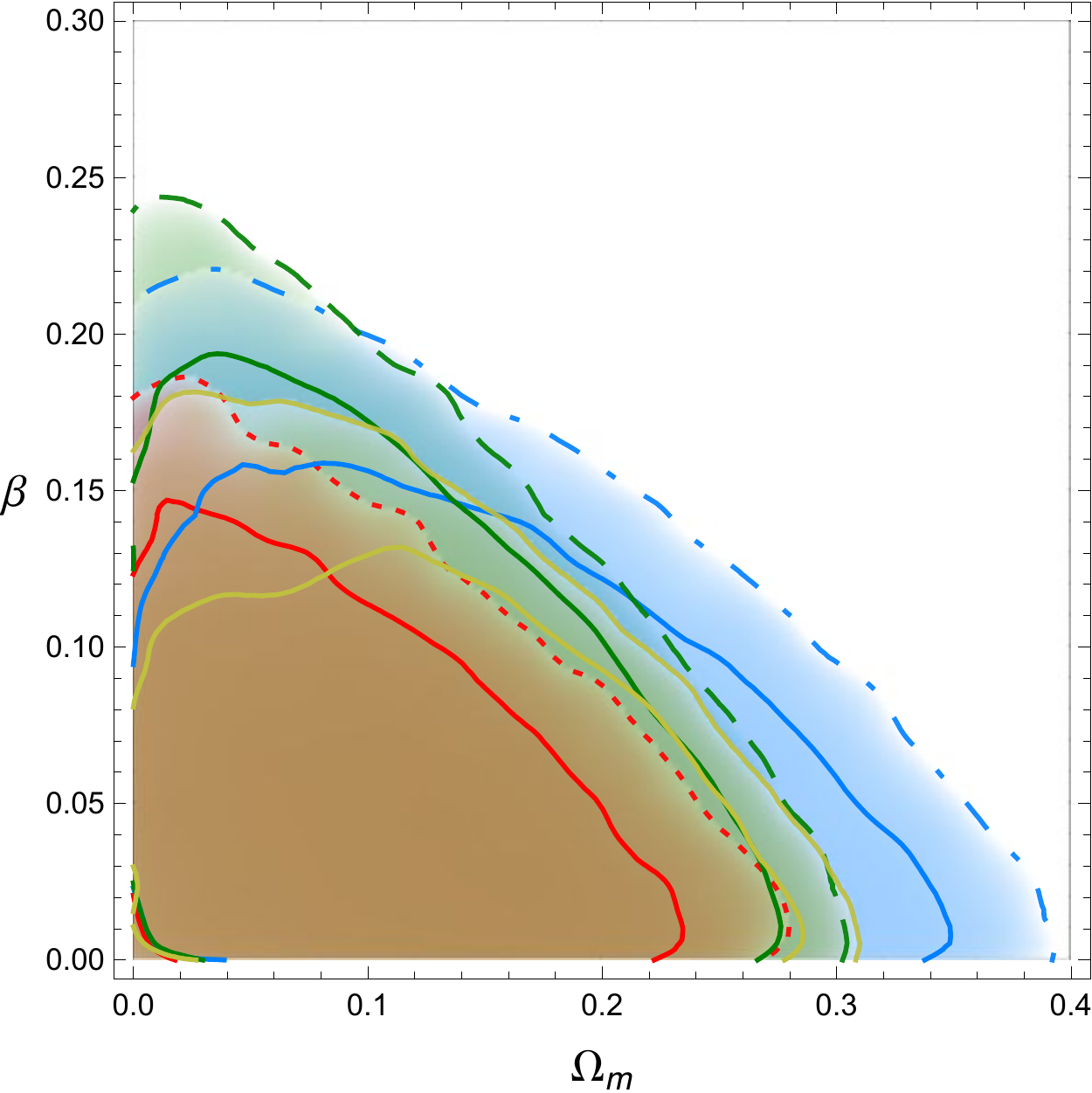}
    \caption{\label{Fig:ContoursQNIOmB} Confidence regions in the $(\Omega_m,\beta)$ plane for the QNs(I) model. The contours correspond to $1\sigma$-$2\sigma$ confidence levels using LGRBs, SNeIa, H(z) data and LGRBs+SNeIa+H(z) data. The code of colours is the same as on the 
    first panel of Figure \ref{Fig:ContoursQNI}. }
\end{figure}

\begin{figure}
\centering
\includegraphics[width=0.45\textwidth]{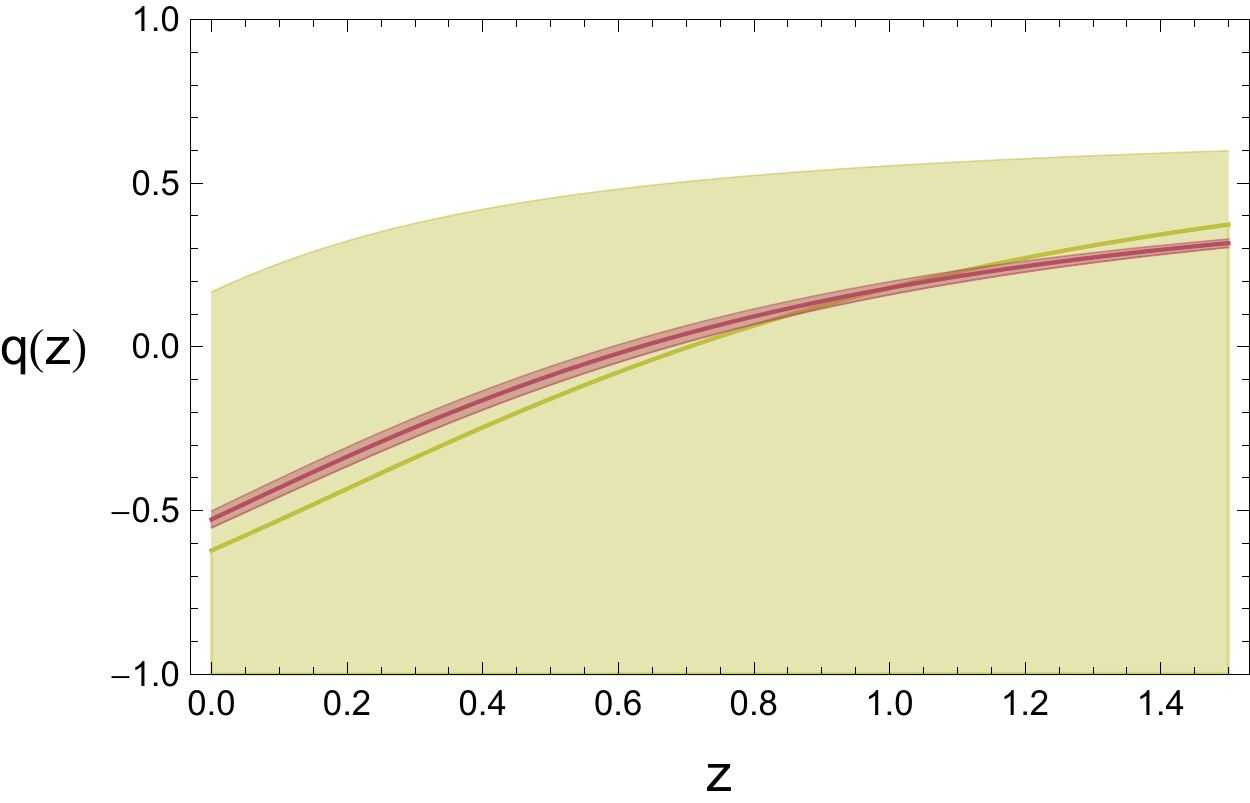}
\caption{\label{Fig:Contoursq2} 
Redshift evolution of the deceleration parameter, $q(z)$, for the QNs(I) model along with $1\sigma$ errors from LGRBs+SNeIa+H(z) data (shaded region in yellow),
and for comparison, the same quantity (shaded tight region in red) as for the $\Lambda$CDM model by assuming $\Omega_m$ representing the baryonic matter fraction with values 
from \citep{PlanckXVI}.}
\label{Fig:qQNI}
\end{figure}

{\renewcommand{\tabcolsep}{1.mm}
{\renewcommand{\arraystretch}{1.5}
\begin{table*}
 \centering
\caption{Median values ($\mu_{1/2}$) for the free parameters of the QNs(I) model and corresponding values for   $q_0$ and  $\Omega_{\Lambda}$ from SNIa, H(z), LGRB data and the combination of all data sets at $1\sigma$ confidence level. The results coming from the combination of all data 
sets obtained by assuming a Gaussian prior on $\Omega_m$ are also shown.}
\label{Table:1}
\begin{tabular}{l|cc|cc|cc||c|c}
\hline \hline
     & \multicolumn{2}{c|}{$\beta$}    & \multicolumn{2}{c|}{$\gamma$}   & \multicolumn{2}{c||}{$\Omega_m$} & $q_0$ & $\Omega_{\Lambda}$   \\
\cline{1-7}
     & $\mu_{1/2}$ & $1\sigma$ & $\mu_{1/2}$ & $1\sigma$ & $\mu_{1/2}$ & $1\sigma$ &       &                     \\
\hline \hline
\small{SNIa}  & $0.070$ & $<0.187$ & $0.124$ & $<0.389$ & $0.116$ & $<0.295$ & $-0.605^{+0.034}_{-0.034}$ &$0.751^{+0.032}_{-0.028} $ \\
\small{H(z)}  & $0.057$ & $<0.112$ & $0.157$ &$<0.468$ & $0.153$ & $<0.370$ & $-0.524^{+0.061}_{-0.061}$&$0.695^{+0.048}_{-0.044} $ \\
\small{LGRBs} & $0.052$ & $<0.175$ & $0.103$ & $<0.349$ & $0.093$ & $<0.251$ & $-0.692^{+0.059}_{-0.054}$ & $0.804^{+0.040}_{-0.042}$ \\
\small{Combination} & $0.048$ & $<0.092$ & $0.141$ & $<0.396$ & $0.135$ & $<0.296$ & $-0.598^{+0.029}_{-0.031}$ & $0.741^{+0.028}_{-0.023}$ \\
\hline
\small{Prior} & $0.060$ & $<0.074$ & $0.236$ & $(0.215,0.350)$ & $0.0490$ & $(0.0481,0.0493)$ & $-0.610^{+0.055}_{-0.011}$ & $0.753^{+0.008}_{-0.049}$ \\
\hline \hline
\end{tabular}
\end{table*}}}

{\renewcommand{\tabcolsep}{1.mm}
{\renewcommand{\arraystretch}{1.5}
\begin{table*}
 \centering
\caption{Median values ($\mu_{1/2}$)  for the free parameters of the QGP(I) model and corresponding values for   $q_0$ and  $\Omega_{m}$ from SNIa, H(z), LGRB data and the combination of all data sets at $1\sigma$ confidence level. The results coming from the combination of all data 
sets obtained by assuming a Gaussian prior on $\Omega_m$ are also shown.}\label{Table:2}
\begin{tabular}{l|cc|cc|c|cc|c}
\hline \hline
     & \multicolumn{2}{c|}{$\beta$}    & \multicolumn{2}{c|}{$\gamma$}    & $q_0$ & $\Omega_{m}$   \\
\cline{1-5}
     & $\mu_{1/2}$ & $1\sigma$ & $\mu_{1/2}$ & $1\sigma$ &  &            \\
\hline \hline
\small{SNIa}  & $0.093$ & $<0.202$ & $0.790$ & $(0.767,0.818)$ &$-0.591^{+0.031}_{-0.029}$ &$<0.297$ \\
\small{H(z)}  &$0.088$ &$<0.162$ & $0.745$& $(0.701,0.777)$&$-0.509^{+0.058}_{-0.054}$ & $(0.088,0.376)$\\
\small{LGRBs} & $0.069$ &$<0.153$ & $0.835$ & $(0.802,0.878)$&$-0.676^{+0.054}_{-0.048}$ &$<0.254$  \\
\small{Combination} & $0.087$ &$<0.171$ & $0.784$ & $(0.761,0.805)$&$-0.580^{+0.027}_{-0.026}$ & $(0.034,0.304)$  \\
\hline
\small{prior} & $0.162$ & $(0.153,0.172)$ & $0.801$ & $(0.791,0.811)$& $-0.607^{+0.019}_{-0.019}$ &$0.049^{+0.001}_{-0.001}$  \\
\hline \hline
\end{tabular}
\end{table*}}}
\begin{figure}
  \centering
  \includegraphics[width=0.45\textwidth]{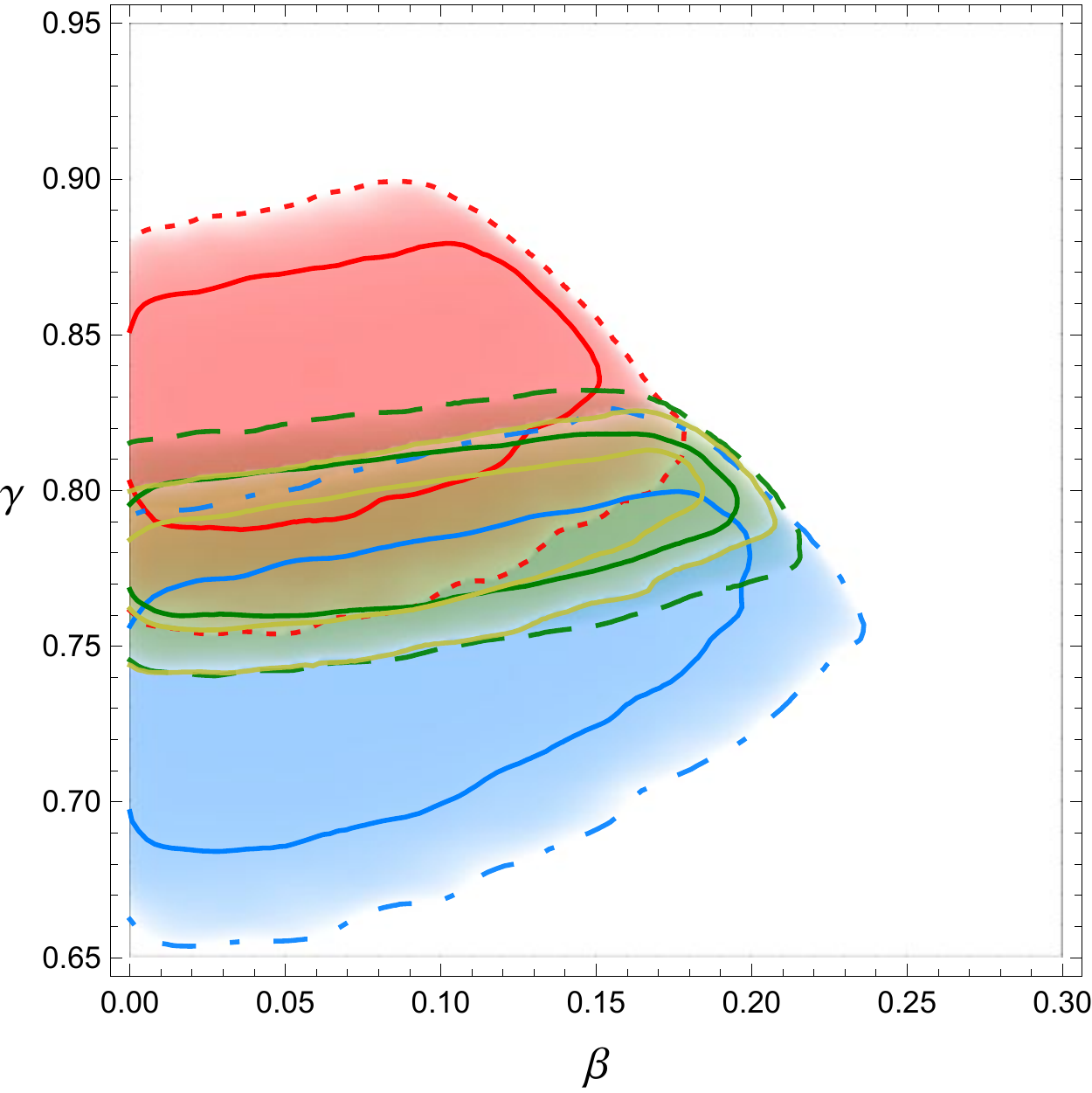}
  \hskip .8cm
    \includegraphics[width=0.45\textwidth]{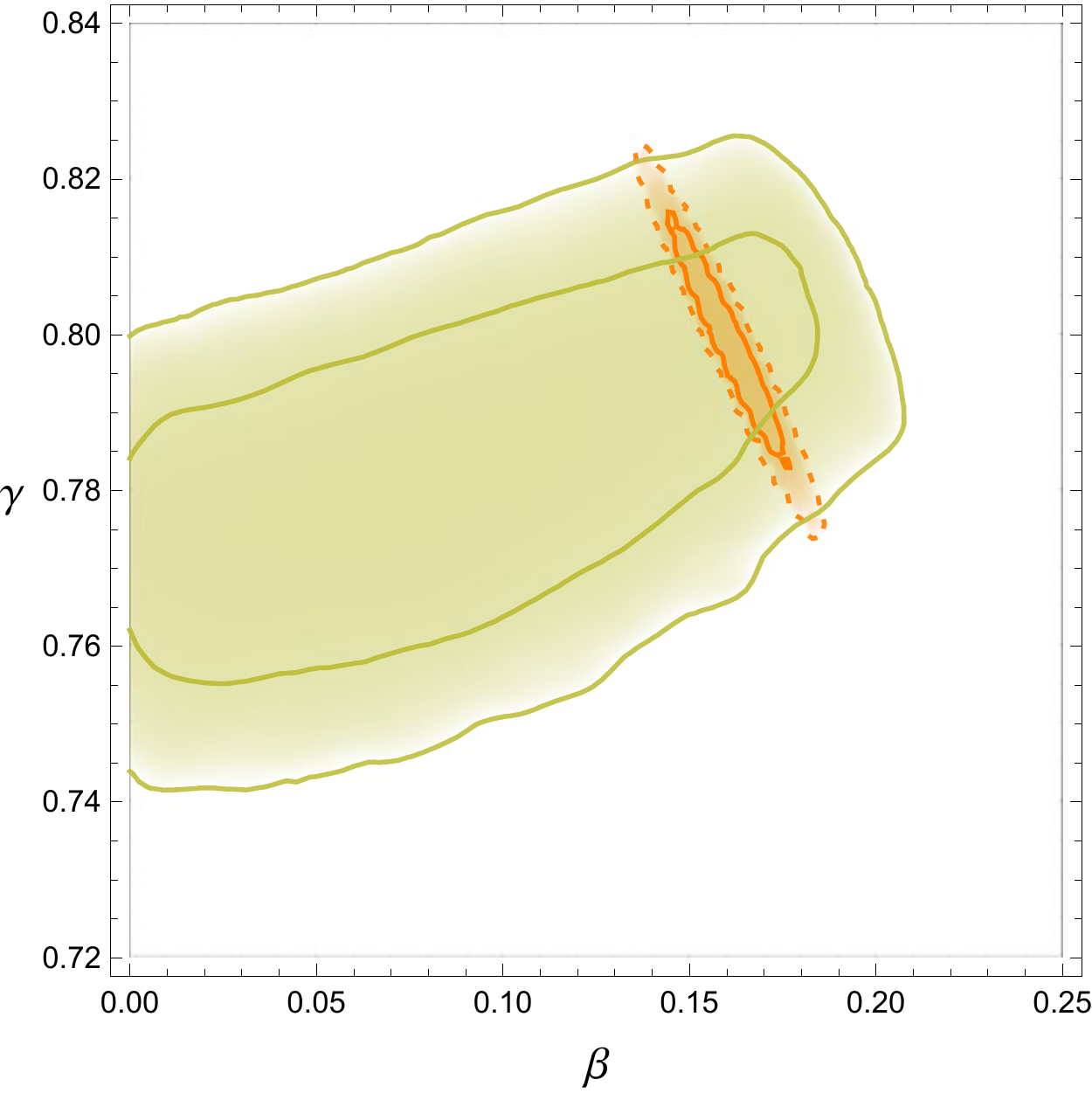}
  \caption{\label{Fig:Contours} Constraints on the free parameters of the QGP(I) model. (\textit{First panel.}) The contours correspond to $1\sigma$-$2\sigma$ confidence levels using LGRBs (region at the top),
  SNeIa (region in the middle), H(z) data (region at the bottom) and LGRBs+SNeIa+H(z) data (tighter region). The code of colours is the same as on the first panel of Figure 
  \ref{Fig:ContoursQNI}. (\textit{Second Panel.}) The contours correspond to $1\sigma$-$2\sigma$ confidence level using LGRBs+SNeIa+H(z) data. The shaded region in orange shows the confidence region that results from the $\chi^2$ analysis by assuming a Gaussian prior on $\Omega_m$ from \citep{PlanckXVI}, while the shaded region 
  in yellow in obtained without assuming a prior knowledge on $\Omega_m$.  }
\end{figure}

\begin{figure}
\centering
\includegraphics[width=0.45\textwidth]{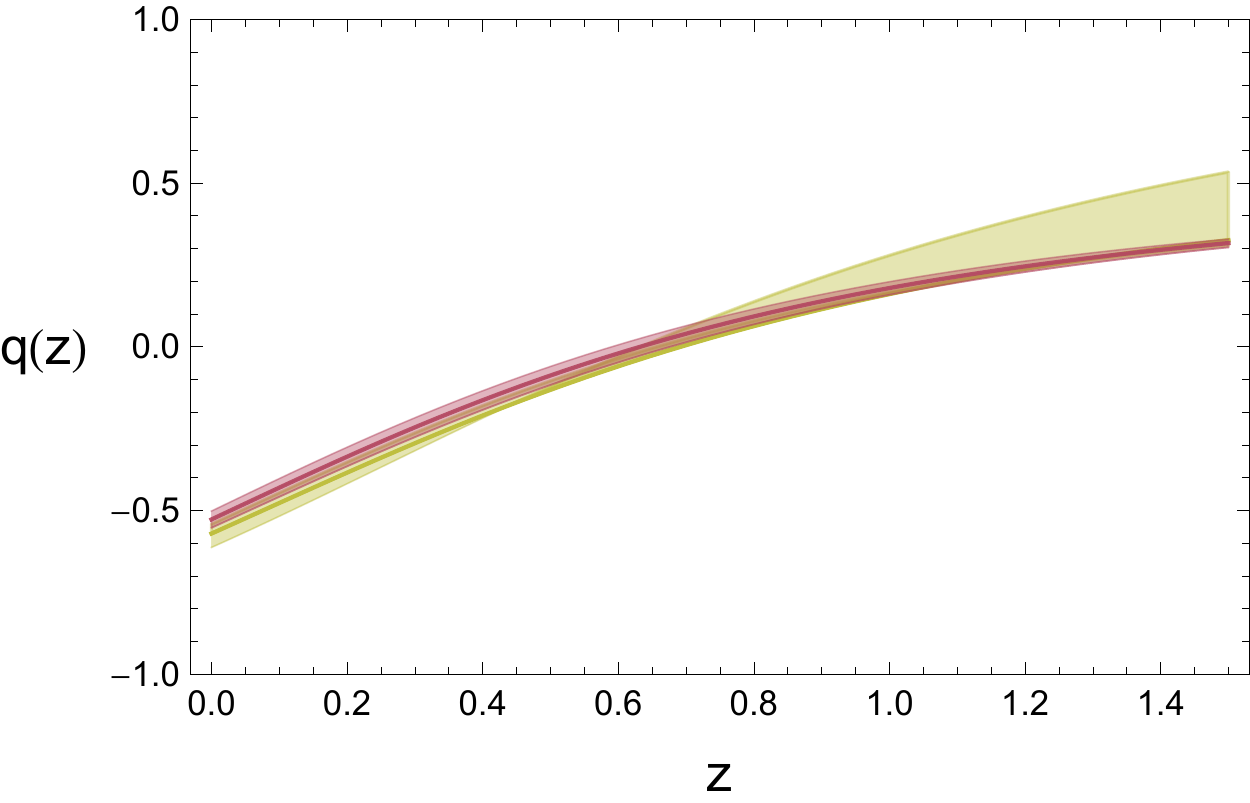}
\caption{\label{Fig:Contoursq} Redshift evolution of the deceleration parameter, $q(z)$, for the QGP(I) model along with $1\sigma$ errors from LGRBs+SNeIa+H(z) data (shaded region in yellow),
and for comparison, the same quantity (shaded tight region in red) as for the $\Lambda$CDM model by assuming $\Omega_m$ from \citep{PlanckXVI}.}
\label{Fig:q}
\end{figure}

In order to constrain the free parameters of the model, we have to 
maximize the posterior probability distribution, which is proportional to the likelihood function $\mathcal{L}(\theta_i) \propto \exp [-\chi^2(\theta_i)/2]$,
and may include some priors and normalization.
We do the sampling of the parameter space using the Markov Chain Monte Carlo Method (MCMC), which is a well-known algorithm widely used for that task, 
obtained following the Bayesian approach. The criteria to decide whether the chain has convergence is its 
main complication, and here, to address this issue, we have followed the prescription developed and described in \cite{Dunkley05}. 
For further insight on MCMC methods, see for example \cite{Berg,MacKay,Neal} and references therein.

The summary of our findings, including an estimation of the deceleration parameter from Eq. (\ref{Eq:q}) and the expected amount of cosmological constant, $\Omega_{\Lambda}$, 
and visible matter, $\Omega_m$, are displayed in Tables \ref{Table:1} and \ref{Table:2}.

Concerning the QN(I) model, we can easily check that their use to explain DM is highly questionable, and statistics does not offer any good conclusion due to the 
high degeneracy between the theoretical parameters. The best-fit values clearly show that $\beta$ is practically consistent with zero, while the role of $\gamma$
as a DM contribution is quite dubious. Even more, from the $1\sigma$ confidence levels, we cannot assure if the DM contribution is resolved by $\Omega_{m}$ 
or by the contribution from the quark fluid $\gamma$. However, when  a Gaussian prior on $\Omega_m$ is assumed (see below), the results clearly show that the model is 
compatible with DM completely determined by $\gamma$, while the $\Omega_m$ term corresponds to the classical baryonic content, see Table \ref{Table:1}. 
Specifically  this prior is $\Omega_m h^2=0.02205\pm 0.00028$ with $H_0= 100 \; h \; {\rm km} {\rm s}^{-1} {\rm Mpc}^{-1}= 67.3~{\rm km} {\rm s}^{-1} {\rm Mpc}^{-1}$, 
as given by \citep{PlanckXVI}.

Figures \ref{Fig:ContoursQNI}, \ref{Fig:ContoursQNIOmG}, and \ref{Fig:ContoursQNIOmB} show the constraints on the free parameters $(\beta, \gamma, \Omega_m)$ for the QN(I) model.  The high degeneracy between the parameters in this case can be noticed. On the other hand, Figure \ref{Fig:qQNI} 
shows the evolution of the deceleration parameter $q$ with $z$. Note that although the prediction for the $q$ parameter is quite similar to the one from the $\Lambda$CDM model, the poor constraints for the free parameters lead to very big errors.

On the other side, the best-fit values of the free parameters of the QGP (I) model from SNeIa data, are $\gamma=0.784^{+0.034}_{-0.051}$ and $\beta<0.202$ at $1\sigma$ confidence level, 
see Table \ref{Table:2}. 
As can be noticed, $\gamma$ is acceptably well constrained, while for $\beta$  only an upper limit can be set, and it turns out to be statistically compatible with zero at the lower-values limit. These results are compatible with 
those obtained from Hubble and LGRB data, although a larger amount of $\gamma$ is allowed from LGRB data 
than from SNeIa or Hubble data.

These results are quite questionable: they seem to imply that a Universe composed by that unknown perfect fluid with the thermodynamical properties inherited by the QGP is 
possible, but the compatibility of $\beta$ with zero makes the model quite equivalent to the classical $\Lambda$CDM model. In this context, the quark fluid could mimic DE 
through the $\gamma$ parameter, but it would be unable to explain DM. This is clearly shown by the values of the $\Omega_{m}$ parameter: it clearly resembles the present 
values for it, i.e., with contribution from baryonic and dark matter, while, from theoretical assumptions, it should be only the baryonic bit.

In order to obtain tighter intervals for the model parameters, we combine all the used data sets and we obtain $\gamma=0.777^{+0.015}_{-0.016}$ and $\beta<0.171$ at $1\sigma$ 
confidence level. Such constraints differ very little from the previous ones, thus corroborating the doubts about the real feasibility of a QGP scenario. In Figure \ref{Fig:Contours}, 
\textit{first panel}, the respective confidence regions can be seen; note that combining all data sets, slightly more stringent confidence regions are achieved (region in solid line).  

On the other hand, our results are also in disagreement with theoretical predictions given in Section 4.1 of \citep{Brilenkov}: the $\beta$ and the $\gamma$ parameters should be 
anti-correlated, while our findings show a positive/null correlation.
Our tests were performed without assuming any prior; in order to validate the previous statement, we performed an additional statistical analysis combining again all data 
sets but assuming a Gaussian prior given by $\Omega_m h^2=0.02205\pm 0.00028$ with $H_0= 100 \; h \; {\rm km} {\rm s}^{-1} {\rm Mpc}^{-1}= 67.3~ {\rm km} {\rm s}^{-1} {\rm Mpc}^{-1}$  
as hinted by \citep{PlanckXVI}. Besides, following the theoretical suggestion from \citep{Brilenkov}, we assume that $\Omega_{m}$ is only baryonic matter.

From this analysis, the best-fit values for $\gamma$ and $\beta$ turn out to be $\beta=0.162^{+0.010}_{-0.009}$ and $\gamma=0.801\pm0.010$. 
Figure \ref{Fig:Contours}, \textit{second panel}, shows the respective confidence region, in which we also draw, for comparison, 
the confidence region obtained from the analysis without any prior and with the combination of all observational data sets. As can be seen, 
the constraints on both parameters are significantly improved, although we have to keep in mind that these impressive results were derived by 
assuming the prior on $\Omega_m$. 

See also Fig. \ref{Fig:q} for a comparison between the evolution of the deceleration parameter $q(z)$  obtained  using LGRBs+SNeIa+H(z) data coming from the QGP(I) model
(shaded region in yellow) and from the $\Lambda$CDM model (shaded tight region in red)  from \citep{PlanckXVI}.

\section{Conclusion}
In short, we have tested the proposals based on the assumption that a small fraction of quarks and gluon survived after an early-universe phase transition in the form of
a quark-gluon 
\textit{nuggets} or as a perfect fluid with which thermodynamical properties received from a QGP. 

With respect to the QNs(I) scenario, our numerical analysis indicates that the role of $\gamma$ as a contribution to DM is only possible when a prior on $\Omega_m$ 
is assumed and that otherwise, it cannot be established which component plays the role of DM. Thus, the claim that Quark Nuggets are candidates for DM have to be taken with 
caution as it us due a result that is strongly dependent on the adopted priors, and, thus, it may  lead to misleading conclusions.

Concerning the QGP(I) scenario, the best-fit values of the model parameters obtained from a statistical analysis with SNeIa, LGRBs and $H(z)$ data allow us to draw the following conclusions:
\begin{itemize}
\item there is not a striking evidence in favour of the QGP(I) model as a way to describe both the accelerated expansion  and to account for the amount expected of DM;  
\item and the assumption of an ad-hoc prior on $\Omega_m$ strongly favours the QGP(I) model, but then this is a weak result as it relies on a strong initial hypothesis (prior).
\end{itemize}
Therefore, all in all, we conclude that the QGP(I) model does not \textit{naturally} explain cosmological dynamics.


\section*{Acknowledgements}
AM acknowledges financial support from Conacyt-M\'exico, through a PhD grant. VS and RL are supported by the Spanish Ministry of Economy and Competitiveness through research projects FIS2010-15492 and Consolider EPI CSD2010-00064, and also by the Basque Government through research project   GIC12/66, and by the University of the Basque Country UPV/EHU under program UFI 11/55.

\bibliography{biblio}

\end{document}